\def\comment#1{}

\documentstyle[12pt]{article}

\skewchar\fivmi='177
\skewchar\sixmi='177
\skewchar\sevmi='177
\skewchar\egtmi='177
\skewchar\ninmi='177
\skewchar\tenmi='177
\skewchar\elvmi='177
\skewchar\twlmi='177
\skewchar\frtnmi='177
\skewchar\svtnmi='177
\skewchar\twtymi='177
\def\@magscale#1{ scaled \magstep #1}
\skewchar\fivsy='60
\skewchar\sixsy='60
\skewchar\sevsy='60
\skewchar\egtsy='60
\skewchar\ninsy='60
\skewchar\tensy='60
\skewchar\elvsy='60
\skewchar\twlsy='60
\skewchar\frtnsy='60
\skewchar\svtnsy='60
\skewchar\twtysy='60

\catcode`@=11
\def\un#1{\relax\ifmmode\@@underline#1\else
  $\@@underline{\hbox{#1}}$\relax\fi}
\catcode`@=12

\def\a{\alpha}
\def\b{\beta}

\def\d{\delta}
\def\e{\epsilon}

\def\g{\gamma}

\def\q{\theta}

\def\s{\sigma}

\def\D{\Delta}

\def\S{\Sigma}

\def\dslash{\not{\hbox{\kern-2pt $\partial$}}}
\def\Dslash{\not{\hbox{\kern-4pt $D$}}}
\def\pslash{\not{\hbox{\kern-2.3pt $p$}}}
 \newtoks\slashfraction
 \slashfraction={.13}
 \def\slash#1{\setbox0\hbox{$ #1 $}
 \setbox0\hbox to \the\slashfraction\wd0{\hss \box0}/\box0 }

\font\ro=cmsy10       
\def\kcr{{\hbox{\ro \char'170}}}    
\def\ktl{{\hbox{\ro \char'170}}}  
\def\ktr{{\hbox{\ro \char'170}}}  
\def\kbl{{\hbox{\ro \char'170}}}  
\def\kbr{{\hbox{\ro \char'170}}}  

\def\plpl{\raise-2pt\hbox{$\raise3pt\hbox{$_+$}\hskip-6.67pt\raise0.0pt
\hbox{$^+$}\hskip 0.01pt$}}
\def\mimi{\raise-2pt\hbox{$\raise3pt\hbox{$_-$}\hskip-6.67pt\raise0.0pt
\hbox{$^-$}\hskip 0.01pt$}}

\def\bo{{\raise.15ex\hbox{\large$\Box$}}}    
\def\pa{\partial}          
\def\TH{{\raise.2ex\hbox{$\displaystyle \bigodot$}\mskip-4.7mu \llap H \;}}
\def\face{{\raise.2ex\hbox{$\displaystyle \bigodot$}\mskip-2.2mu \llap
{$\ddot \smile$}}}          
\def\dg{\sp\dagger}          

\def\sp#1{{}^{#1}}        
\def\Tilde#1{\widetilde{#1}}     
\def\Hat#1{\widehat{#1}}      
\def\Bar#1{\overline{#1}}      
\def\leftrightarrowfill{$\mathsurround=0pt \mathord\leftarrow \mkern-6mu
  \cleaders\hbox{$\mkern-2mu \mathord- \mkern-2mu$}\hfill
  \mkern-6mu \mathord\rightarrow$}
\def\dvec#1{\vbox{\ialign{##\crcr
  \leftrightarrowfill\crcr\noalign{\kern-1pt\nointerlineskip}
  $\hfil\displaystyle{#1}\hfil$\crcr}}}   

\def\frac#1#2{{\textstyle{#1\over\vphantom2\smash{\raise.20ex
  \hbox{$\scriptstyle{#2}$}}}}}     
\def\sfrac#1#2{{\vphantom1\smash{\lower.5ex\hbox{\small$#1$}}\over
  \vphantom1\smash{\raise.4ex\hbox{\small$#2$}}}} 
\def\bfrac#1#2{{\vphantom1\smash{\lower.5ex\hbox{$#1$}}\over
  \vphantom1\smash{\raise.3ex\hbox{$#2$}}}}  
\def\afrac#1#2{{\vphantom1\smash{\lower.5ex\hbox{$#1$}}\over#2}} 
\def\partder#1#2{{\partial #1\over\partial #2}} 

\newskip\humongous \humongous=0pt plus 1000pt minus 1000pt
\def\caja{\mathsurround=0pt}
\def\eqalign#1{\,\vcenter{\openup2\jot \caja
  \ialign{\strut \hfil$\displaystyle{##}$&$
  \displaystyle{{}##}$\hfil\crcr#1\crcr}}\,}
\newif\ifdtup

\def\ref#1{$\sp{#1)}$}

\parskip=\medskipamount     
\thicklines       

\def\oldheadpic{        
  \setlength{\unitlength}{.4mm}
  \thinlines
  \par
  \begin{picture}(349,16)
  \put(325,16){\line(1,0){4}}
  \put(330,16){\line(1,0){4}}
  \put(340,16){\line(1,0){4}}
  \put(335,0){\line(1,0){4}}
  \put(340,0){\line(1,0){4}}
  \put(345,0){\line(1,0){4}}
  \put(329,0){\line(0,1){16}}
  \put(330,0){\line(0,1){16}}
  \put(339,0){\line(0,1){16}}
  \put(340,0){\line(0,1){16}}
  \put(344,0){\line(0,1){16}}
  \put(345,0){\line(0,1){16}}
  \put(329,16){\oval(8,32)[bl]}
  \put(330,16){\oval(8,32)[br]}
  \put(339,0){\oval(8,32)[tl]}
  \put(345,0){\oval(8,32)[tr]}
  \end{picture}
  \par
  \thicklines
  \vskip.2in}
\def\oldtitle#1#2#3#4{\oldheadpic\begin{center}\vglue.5in{\large\bf
   #1}\\[.6in]
  {#2}\\[.1in] {\it Department of Physics and Astronomy}\\
  {\it University of Maryland, College Park, MD 20742}\\[.6in]
  Physics Publication \#{#3}\\ {#4}\\[1.5in] {\bf ABSTRACT}\\[.1in]
  \end{center} \begin{quotation}}     
\def\oldTitle#1#2#3#4#5#6#7{\oldheadpic\begin{center} \vglue .4in
  {\large\bf #1}\\[.4in]
  {#2}\\[.1in] {\it Department of Physics and Astronomy}\\
  {\it University of Maryland, College Park, MD 20742}\\[.1in]
  {#3}\\[.1in] {\it {#4}}\\ {\it {#5}}\\[.4in]
  Physics Publication \#{#6}\\ {#7}\\[.5in] {\bf ABSTRACT}\\[.1in]
  \end{center} \begin{quotation}}     
\def\border{           
  \setlength{\unitlength}{1mm}
  \newcount\xco
  \newcount\yco
  \xco=-21
  \yco=12
  \begin{picture}(140,0)
  \put(\xco,\yco){$\ktl$}
  \advance\yco by-1
  {\loop
  \put(\xco,\yco){$\kcr$}
  \advance\yco by-2
  \ifnum\yco>-240
  \repeat
  \put(\xco,\yco){$\kbl$}}
  \xco=158
  \yco=12
  \put(\xco,\yco){$\ktr$}
  \advance\yco by-1
  {\loop
  \put(\xco,\yco){$\kcr$}
  \advance\yco by-2
  \ifnum\yco>-240
  \repeat
  \put(\xco,\yco){$\kbr$}}
  \put(-20,13){\tiny University of Maryland Elementary Particle
Physics University of Maryland Elementary Particle Physics University of
Maryland Elementary Particle Physics}
  \put(-20,-241.5){\tiny University of Maryland Elementary
Particle Physics University of Maryland Elementary Particle Physics
University of Maryland Elementary Particle Physics}
  \end{picture}
  \par\vskip-8mm}
\def\bordero{           
  \setlength{\unitlength}{1mm}
  \newcount\xco
  \newcount\yco
  \xco=-31
  \yco=12
  \begin{picture}(140,0)
  \put(\xco,\yco){$\ktl$}
  \advance\yco by-1
  {\loop
  \put(\xco,\yco){$\kclr}
  \advance\yco by-2
  \ifnum\yco>-240
  \repeat
  \put(\xco,\yco){$\kbl$}}
  \xco=151
  \yco=12
  \put(\xco,\yco){$\ktr$}
  \advance\yco by-1
  {\loop
  \put(\xco,\yco){$\kcr$}
  \advance\yco by-2
  \ifnum\yco>-240
  \repeat
  \put(\xco,\yco){$\kbr$}}
  \put(-20,12){\ooo
bacdefghidfghghdhededbihdgdfdfhhdheidhdhebaaahjhhdahba

hgdedge
 hgfdiehhgdigicba}
  \put(-20,-241.5){\ooo
ababaighefdbfghgeahgdfgafagihdidihiidhiagfedhadbfd

ecdcdfa
 gdcbhaddhbgfchbgfdacfediacbabab}
  \end{picture}
  \par\vskip-8mm}
\def\headpic{           
  \indent
  \setlength{\unitlength}{.4mm}
  \thinlines
  \par
  \begin{picture}(29,16)
  \put(165,16){\line(1,0){4}}
  \put(170,16){\line(1,0){4}}
  \put(180,16){\line(1,0){4}}
  \put(175,0){\line(1,0){4}}
  \put(180,0){\line(1,0){4}}
  \put(185,0){\line(1,0){4}}
  \put(169,0){\line(0,1){16}}
  \put(170,0){\line(0,1){16}}
  \put(179,0){\line(0,1){16}}
  \put(180,0){\line(0,1){16}}
  \put(184,0){\line(0,1){16}}
  \put(185,0){\line(0,1){16}}
  \put(169,16){\oval(8,32)[bl]}
  \put(170,16){\oval(8,32)[br]}
  \put(179,0){\oval(8,32)[tl]}
  \put(185,0){\oval(8,32)[tr]}
  \end{picture}
  \par\vskip-6.5mm
  \thicklines}
\def\title#1#2#3#4{\border\headpic {\hbox to\hsize{#4 \hfill UMDEPP #3}}\par
  \begin{center} \vglue .5in {\large\bf #1}\\[.6in]
  {#2}\\[.1in] {\it Department of Physics and Astronomy}\\
  {\it University of Maryland, College Park, MD 20742}\\[1.5in]
  {\bf ABSTRACT}\\[.1in] \end{center} \begin{quotation}} 
\def\Title#1#2#3#4#5#6#7{\border\headpic
  {\hbox to\hsize{#7 \hfill UMDEPP #6}}\par
  \begin{center} \vglue .4in {\large\bf #1}\\[.4in]
  {#2}\\[.1in] {\it Department of Physics and Astronomy}\\
  {\it University of Maryland, College Park, MD 20742}\\[.1in]
  {#3}\\[.1in] {\it {#4}}\\ {\it {#5}}\\[.5in] {\bf ABSTRACT}\\[.1in]
  \end{center} \begin{quotation}}     
\def\endtitle{\end{quotation}\newpage}     

\def\sect#1{\bigskip\medskip \goodbreak \noindent{\bf {#1}} \nobreak \medskip}

\def\ad{{\kern0.5pt
     \alpha \kern-5.05pt \raise5.8pt\hbox{$\textstyle.$}\kern
0.5pt}}

\def\bd{{\kern0.5pt
     \beta \kern-5.05pt \raise5.8pt\hbox{$\textstyle.$}\kern
0.5pt}}

\def\qd{{\kern0.5pt
     q \kern-5.05pt \raise5.8pt\hbox{$\textstyle.$}\kern
0.5pt}}
\def\Dot#1{{\kern0.5pt
     {#1} \kern-5.05pt \raise5.8pt\hbox{$\textstyle.$}\kern
0.5pt}}

\begin{document}

\def\gfrac#1#2{\frac {\scriptstyle{#1}}
  {\mbox{\raisebox{-.6ex}{$\scriptstyle{#2}$}}}}
\def\gg{{\hbox{\sc g}}}
\border\headpic {\hbox to\hsize{December 1996 \hfill {UMDEPP 97-27}}}
{\hbox to\hsize{ \hfill {BRX-TH-397}}}
{\hbox to\hsize{ \hfill {WATPHYS-TH96/14}}}
{\hbox to\hsize{ \hfill {ITP-SB-96-74}}}
\par
\setlength{\oddsidemargin}{0.3in}
\setlength{\evensidemargin}{-0.3in}
\begin{center}
\vglue .04in
{\large\bf N = 1 Supersymmetric Extension of the\\
QCD Effective Action }
\\[.1in]

S. James Gates, Jr. \footnote{ gates@umdhep.umd.edu.
Supported in part by NSF Grant PHY-96-43219 and Nato Grant \newline ${~~\,~~}$
CRG-93-0789} \\
{\it Department of Physics,
University of Maryland at College Park \\
College Park, MD 20742-4111, USA. } \\[.08in]
M.~T.~Grisaru\footnote{grisaru@binah.cc.brandeis.edu. Supported in part by NSF
Grant PHY-92-22318}\\
{\it Physics Department,
Brandeis University\\
Waltham, MA 02254-9110, USA.}\\[.08in]
Marcia~E.~Knutt-Wehlau\footnote{marcia@avatar.uwaterloo.ca. John Charles
Polanyi Fellow. Supported in part by NSERC of \newline ${~~\,~~}$
Canada and a NSERC Postdoctoral Fellowship}\\
{\it Physics Department,
University of Waterloo\\
Waterloo, Ont.
Canada N2L 3G1.}\\[.08in]
M.\ Ro\v cek\footnote{rocek@insti.physics.sunysb.edu. Supported in part by NSF
Grant PHY-93-09888}
\\
{\it Institute for Theoretical Physics,
State University of New York\\
Stony Brook, NY 11794-3840, USA.}\\[.08in]
Oleg A. Soloviev\footnote{soloviev@v1.ph.qmw.ac.uk}
\\
{\it Physics Department, Queen Mary and Westfield College\\
Mile End Road, London E1 4NS
United Kingdom.}
\\ [.4in]

{\bf ABSTRACT}\\[.002in]
\end{center}
\begin{quotation}
{We present a new 4D, N = 1 supersymmetric nonlinear $\s$-model using 
complex linear and chiral superfields that generalizes the massless limit 
of the QCD effective action of Gasser and Leutwyler.}
\endtitle

\sect{I. Introduction}

It is well known that the QCD low-energy effective action is accurately
described by a (broken) $SU(3)_L\otimes SU(3)_R$ chiral model encoding
the interactions of the lightest flavor-$SU(3)$ meson octet \cite{lit}. A
systematic expansion in momenta has been given by Gasser and Leutwyler
\cite{gasser}. Though we know that supersymmetry is not a feature of
low-energy phenomenology, a great deal of qualitative information about
field theories in general has been learned from the study of supersymmetric
theories and it is of interest to study the low-energy effective actions
for N = 1 supersymmetric QCD.

Earlier studies have been formulated solely in terms of chiral superfields
\cite{attempts}. Recently, one of us \cite{jim} has proposed a new class of
models to describe supersymmetric extensions of the low-energy QCD effective
action. These models use both chiral (C) and nonminimal (N), complex linear,
superfields to describe scalar multiplets, and split the physical component
fields between them. We refer to them as CNM models. The splitting is
essentially heterodexterous with ``right-handed'' matter contained in
C-superfields and ``left-handed'' matter contained in N-superfields.
The spinor superpartners of the pions in the CNM models are most naturally 
described by Dirac fields.  Geometrically, the chiral superfields are 
coordinates in the cotangent bundle of the $\sigma$-model manifold  considered 
as a base manifold, while the complex linear superfields coordinatize the 
fibers.

It is always possible, using suitable duality transformations, to
{\it {formally}} replace the nonminimal superfields by chiral superfields.
To leading order in (spinor) derivatives the resulting theory looks like
the proposal of Rohm and Nemeschansky \cite{attempts}.  It differs by
higher derivative terms and by having twice as many fields, which enables
the spinor superpartners of the pions to also be Dirac fields in the dual
version of the CNM model. Furthermore, the CNM formulation allows an
{\it {explicit}} mapping from the bosonic effective action to the N = 1 
supersymmetric effective action, and exhibits some interesting geometrical 
features \cite{jim}.

In the simplest case, we are interested in constructing a 4D, N = 1
superfield action with the property that if only the pion octet is retained,
the manifestly supersymmetric action should come as close as possible to
being in agreement with the Gasser-Leutwyler parametrization. An important
ingredient in this is the explicit realization of chiral $SU(3)_L\otimes 
SU(3)_R$ symmetry in terms of superfields. To simplify matters further, we 
work in the limit where all masses vanish and all spin-1 fields are set 
to zero.  What emerges is a simple form that is the topic of this Letter.

\newpage

\sect{II. The Massless Limit of the Gasser-Leutwyler Action}

We recall that the pion octet can be introduced as an element of the $SU(3)$
algebra in the form,
$$ \frac 1{f_{\pi}} \Pi ~\equiv~
\frac 1{ f_{\pi}} \Pi^i t_i ~=~ \frac 1{ f_{\pi}} \left(\begin{array}{ccc}
{}~\frac{\pi^0}{\sqrt 2} ~+~ \frac{\eta}{\sqrt 6} & ~~\pi^+ & ~~K^+ \\
{}~\pi^- & ~~-\, \frac{\pi^0}{\sqrt 2} ~+~ \frac{\eta}{\sqrt 6} & ~~K^0\\
{}~K^- & ~~{\Bar K}^0 & ~~ - \eta \sqrt {\frac 23} \\
\end{array}\right) ~~~~.
\eqno(1) $$
Here $t_1 , \,..., t_8$ are the Gell-Mann $SU(3)$ matrices and $f_{\pi}$ is
the pion decay constant. Exponentiation of the Lie algebraic element in (1)
via the definition $U \equiv \exp [ i \frac 1{ f_{\pi}} \Pi^i t_i ]$ then leads
to elements in the $SU(3)$ group. Other useful quantities are the Maurer-Cartan
forms $R_m {}^i$ and $L_m {}^i$ introduced via
$$
U^{-1} \pa_{\underline a} U ~=~ i {f_{\pi}}^{-1} ( \, \pa_{\underline a} \Pi^m
\,)~ R_m {}^i (\Pi) \, t_i ~~~~,~~~~ (\, \pa_{\underline a} U \,) U^{-1} ~=~
i {f_{\pi}}^{-1} ( \, \pa_{\underline a} \Pi^m \,) ~L_m {}^i
(\Pi) \, t_i ~~~~.
\eqno(2) $$
(We are using underlined letters to denote space-time vector indices, for later
use in the supersymmetric case.)

These definitions allow $R_m {}^i (\Pi)$ and $L_m {}^i (\Pi)$ to be calculated
as power series in $\Pi^i$ from
$$\eqalign{
R_m {}^i (\Pi) &\equiv ~ (C_2)^{-1} {\rm {Tr}} \Big[\, t^i \Big(
\frac { 1 ~-~ e^{-\D} }{\D} \Big) t_m \Big] ~~~, \cr
L_m {}^i (\Pi) &\equiv ~ (C_2)^{-1} {\rm {Tr}} \Big[\, t^i \Big(
\frac { e^{\D} ~-~ 1 }{\D} \Big) t_m \Big] ~~~, }
\eqno(3) $$
where $\D t_m \equiv i {f_{\pi}}^{-1} [ \Pi \, , \, t_m ]$, $\D^2 t_m
= \D \D t_m$, etc. and the constant $C_2$ is determined so that $ L_m {}^i
(0) = R_m {}^i (0) = \d_m {}^i$.

As a consequence, the actions (which occur in \cite{gasser})
$$ {\cal S}_{\s} ~=~ \frac {f_{\pi}^2}{2 C_2} \, \int d^4 x ~
{\rm {Tr}} [\, ( \pa^{\un a} U^{-1} \,) ~ (\pa_{\un a} U \,) \,]
{}~~~~,
\eqno(4) $$
$$ {\cal S}_{1} ~=~ L_1 \, \int d^4 x ~
\Big( ~{\rm {Tr}} [\, ( \pa^{\un a} U^{-1} \,) ~ (\pa_{\un a} U \,) \,]
{}~ \Big)^2
{}~~~~,
\eqno(5) $$
$$ {\cal S}_{2} ~=~ L_2 \, \int d^4 x ~
{\rm {Tr}} [\, ( \pa_{\un a} U^{-1} \,) ~ (\pa_{\un b} U \,) \,] \,
{\rm {Tr}} [\, ( \pa^{\un a} U^{-1} \,) ~ (\pa^{\un b} U \,) \,]
{}~~~~,
\eqno(6) $$
$$ {\cal S}_{3} ~=~ L_3 \, \int d^4 x ~
{\rm {Tr}} [\, ( \pa_{\un a} U^{-1} \,) ~ (\pa_{\un b} U \,) \,
( \pa^{\un a} U^{-1} \,) ~ (\pa^{\un b} U \,) \,]
{}~~~~,
\eqno(7) $$
can be re-expressed in terms of the Maurer-Cartan forms.
In fact there is some redundancy in (5) and (6), and also (7)
is not written in a manner that is most convenient to discuss
the possible supersymmetric extension. To sort this out, it
is convenient to introduce irreducible projection operators,
$$ \eqalign{ {~~}
{P}^{(0) ~ \un a \, \un b \, \un c \, \un d} &\equiv~ \frac 14
\eta^{\un a \, \un b} ~ \eta^{\un c \, \un d} ~~~~~~~, ~~~~~~~
{P}^{(1 ) ~ \un a \, \un b \, \un c \, \un d} ~\equiv~
\frac 12 \, \Big[~ \eta^{\un a \, \un c} ~ \eta^{\un d \, \un b} ~-~
\eta^{\un a \, \un d} ~ \eta^{\un c \, \un b} ~\Big]
 ~~~, \cr
{P}^{(2) ~ \un a \, \un b \, \un c \, \un d} &\equiv~
\frac 12 \, \Big[~ \eta^{\un a \, \un c} ~ \eta^{\un d \, \un b} ~+~
\eta^{\un a \, \un d} ~ \eta^{\un c \, \un b} ~-~ \frac 12
\eta^{\un a \, \un b} ~ \eta^{\un c \, \un d} ~\Big]
 ~~~. } \eqno(8) $$
It follows from the completeness of these operators that we
may write
$$
{\cal S}_{2} ~=~ {\cal S}_{2}^{(0)} ~+~ {\cal S}_{2}^{(1)} ~+~
{\cal S}_{2}^{(2)} ~~~, \eqno(9) $$
$$
{\cal S}_{3} ~=~ {\cal S}_{3}^{(0)} ~+~ {\cal S}_{3}^{(1)} ~+~
{\cal S}_{3}^{(2)} ~~~, \eqno(10) $$
where
$$
{\cal S}_{2}^{(i)} ~\equiv~ L_2 {}^{(i)} \, \int d^4 x ~
{P}^{(i) ~ \un a \, \un b \, \un c \, \un d} ~
{\rm {Tr}} [\, ( \pa_{\un a} U^{-1} \,) ~ (\pa_{\un b} U \,) \,] \,
{\rm {Tr}} [\, ( \pa_{\un c} U^{-1} \,) ~ (\pa_{\un d} U \,) \,]
{}~~~~,
\eqno(11) $$
$$
{\cal S}_{3}^{(i)} ~\equiv~ L_3 {}^{(i)} \, \int d^4 x ~
{P}^{(i) ~ \un a \, \un b \, \un c \, \un d} ~
{\rm {Tr}} [\, ( \pa_{\un a} U^{-1} \,) ~ (\pa_{\un b} U \,) \,
 ( \pa_{\un c} U^{-1} \,) ~ (\pa_{\un d} U \,) \,]
{}~~~~.
\eqno(12) $$
Clearly ${\cal S}_1$ and ${\cal S}_{2}^{(0)}$ have the same functional
form so that one of them is redundant. It is therefore consistent to set
${\cal S}_1$ to zero and use ${\cal S}_{2}^{(0)}$ to parametrize its
contribution to physical processes. In the work by Gasser and Leutwyler,
{\it a} {\it {priori}} assumptions that $L_2^{(0)}= L_1 + L_2$, $L_2^{(1)}
= L_2$, $L_2^{(2)}= L_2$ and $L_3^{(i)} = L_3$ were made. Finally, we note
that ${\cal S}_{3}^{(1)}$ describes the familiar ``Skyrme'' term \cite{skyrme},
whose supersymmetric generalization was attempted in ref. \cite{berg}.

Thus to this order, the low-energy QCD effective action may be parametrized
in the form
$$
{\cal S}_{eff} (QCD) ~=~ {\cal S}_{\s} ~+~ \sum_{i = 0}^2
{}~ \Big[ ~ {\cal S}_{2}^{(i)} ~+~ {\cal S}_{3}^{(i)} ~ \Big]
{}~+~ {\cal S}_{WZNW} ~~~~,
\eqno(13) $$
where we have dropped the assumptions mentioned above and have added
in the Wess-Zumino-Novikov-Witten term as well. As shown by Witten
\cite{witten}, it is most convenient to define an extended group element
$\Hat U$. Thus we define $\Hat U \equiv \exp [\, i y f_{\pi}^{-1} \Pi
\,]$ and in terms of the extended group element, the WZNW term is given
by
$$ \eqalign{ {~~~~~~~~}
{\cal S}_{WZNW} &=~ - i N_C \, [ \, 2 {\cdot} 5! \, ]^{-1}
\int d^4 x \, \int_0^1 d y ~ {\rm {Tr}} \Big[ \, ( {\Hat U}^{-1}
\pa_y {\Hat U} \,) ~ {\Hat {\cal W}}_4 \, \Big] ~~~~, \cr
{\Hat {\cal W}}_4 &=~ \e^{{\un a}{\un b}{\un c}{\un d}} \,
(\pa_{\un a} {\Hat U} ^{-1} \,) \, (\pa_{\un b} {\Hat U} \,) \,
(\pa_{\un c} {\Hat U}^{-1} \,) \, (\pa_{\un d} {\Hat U}
\,) ~~~~. }
\eqno(14) $$

We may rewrite the higher derivative terms in (13) in terms of space-time
derivatives of the pion fields, using
$$ \eqalign{
\pa_{\un a} U &=~ \big( \partder U{\Pi^i} \big) ~ \big( \pa_{\un a} \Pi^i
\big) ~ \equiv \big( \pa_i U \big) ~ \big( \pa_{\un a} \Pi^i
\big){}~~~ , \cr
{~~~~~}
\big( \pa_{\un a} U^{-1} \big) \, \big( \pa_{\un b} U \big) &=~
\big( \pa_i U^{-1} \big) \, \big( \pa_j U \big) ~
\big( \pa_{\un a} \Pi^i\big)\big( \pa_{\un b} \Pi^j\big)
{}~\equiv~ {\cal Z}_{i \, j} \big( \pa_{\un a} \Pi^i\big)\big( \pa_{\un b}
\Pi^j\big) ~~~. }
\eqno(15) $$
\newpage \noindent
Here we see the appearance of factors of the form $\pa_{\un a} \Pi^i$,
``the pullback'' from the group manifold to the
spacetime manifold.

We note in (14) that the pullbacks with our definitions are $y$-independent, thus:
$$ \eqalign{ {~~~~~~}
{\cal S}_{WZNW} &=~ - i N_C \, [ \, 2 {\cdot} 5! \, ]^{-1}
\int d^4 x \e^{{\un a_1}{\un a_2}{\un a_3}{\un a_4}} \,
\b_{i_1i_2i_3i_4} (\Pi) ~ \Big( \prod_{j=1}^4 {\pa_{\un a_j} \Pi^{i_j}}
\Big)  ~~~~, \cr
\b_{ijkl} (\Pi) \, &\equiv \, \int_0^1 d y ~
\, {\rm {Tr}} \Big[ \, ( {\Hat U}^{-1} \pa_y {\Hat U} \,) \, (\pa_{i}
{\Hat U} ^{-1} \,) \, (\pa_{j} {\Hat U} \,) \, (\pa_{k} {\Hat U}^{-1}
\,) \, (\pa_{l} {\Hat U} \,) \,\Big] ~~~. }\eqno(16) $$
It is clear that there exist polynomials ${\cal
J}^{A\, 2}_{i_1 \, i_2 \, i_3 \, i_4 }$ and ${\cal J}^{A\, 3}_{i_1 \,
i_2 \, i_3 \, i_4 }$ which are quadratic in ${\cal Z}_{i \, j}$ such that
$$ \eqalign{ {~~~~~}
{\cal S}_2 &=~
\int d^4 x ~ \sum_{A = 0}^2 L_2 {}^{(A)} \, P^{(A) \, {\un a}_{1} \, ...
{\un a}_4} ~ {{Tr}} \Big[ \, {\cal J}^{A\, 2}_{i_1 \, i_2 \, i_3 \, i_4 }
( {{\cal Z}} \,) ~ \Big] \, \Big( \prod_{j=1}^4 {\pa_{\un a_j} \Pi^{i_j}}
\Big) ~~~, }
\eqno(17) $$
$$ \eqalign{ {~~~~~}
{\cal S}_3 &=~
\int d^4 x ~ \sum_{A = 0}^2 L_3 {}^{(A)} \, P^{(A) \, {\un a}_{1} \, ...
{\un a}_4} ~ {{Tr}} \Big[ \, {\cal J}^{A \, 3}_{i_1 \, i_2 \, i_3 \, i_4 }
( {{\cal Z}} \,) ~ \Big] \, \Big( \prod_{j=1}^4 {\pa_{\un a_j} \Pi^{i_j}}
\Big)~~~, }
\eqno(18) $$
where $P^{(0) \, {\un a}_{1} \, ... {\un a}_4}$, $P^{(1) \, {\un a}_{1}
\, ... {\un a}_4}$ and $P^{(2) \, {\un a}_{1} \, ... {\un a}_4}$ were
defined in (8).  We note that higher order (in momentum) terms of the QCD
effective action have the same general form.

\sect{III. The CNM Model on the Group Manifold}

It is our  goal to describe an N = 1 supersymmetric theory
that contains an appropriate generalization of each of the terms that
appear in (13).   In ref. \cite{jim}, a new proposal was made as to how
manifestly 4D, N = 1 supersymmetric formulations containing (16), (17)
and (18) might be obtained. The key idea is that all the spin-0
and spin-1/2 fields do {\it {not}} occur as the components of chiral
scalar superfields (i.e. Wess-Zumino multiplets) as in the standard
orthodoxy \cite{moha} for such constructions. Instead it was proposed
that the right-handed components of Dirac fields be embedded into
Wess-Zumino chiral scalar superfields $\Phi$, $\bar{D}_{\dot{\alpha}}
\Phi = 0$ and the left-handed components of Dirac fields be embedded
into ``non-minimal'' scalar multiplets \cite{nonmin} (i.e. complex
linear superfields) $\S$, $\bar{D}^2 \S=0$.
Supersymmetrizing the nonlinear $\sigma$-model necessarily leads to
a ``parity-doubling" of the basic degrees of freedom: supersymmetry
requires a scalar partner for every pseudoscalar.  Here, we double once
more by introducing the complex linear fields $\Sigma$.

The main benefit of our approach is twofold: (a) the equations of motion of
auxiliary fields are {\it {always}} purely algebraic so that the auxiliary
fields, which do not propagate at the free lagrangian level, remain
non-propagating  -- we refer to this feature as {\em auxiliary freedom};
and (b.) there is an obvious holomorphy condition that we can impose.
(In the final section we will find ways of rewriting our action that makes
this holomorphy condition less natural.  However, dropping holomorphy
{\it {may}} lead to propagation of auxiliary fields.)

Let ${\Phi}^{\rm I}$ be a set of 4D, N = 1 chiral superfields.  Define
chiral superfield group elements by
$$
{U} (\Phi) ~\equiv~ \exp \Big[ {{ \, {\Phi}^{\rm I}
t_{\rm I}}\over{~ f_{\pi}\, cos (\g_{\rm S}) ~}} \Big] ~~~~~,
\eqno(19) $$
where $t_{\rm I}$ denote the hermitian matrix generators of some compact
Lie algebra. Here $\g_{\rm S}$ is a mixing angle (see below) which is
restricted by the form of the supersymmetric WZNW action to satisfy the
condition $sin(2 \g_{\rm S} ) \neq 0$ \cite{jim}.
Due to the complex nature of chiral superfields, we are working in the
complexification of the group with generators $t_{\rm I}$, and
$$
{ U}\dg ~\ne ~ { U}^{-1}  ~~~~.
\eqno(20) $$
More explicitly we may write
$$
{U} (\Phi) = \exp\Big[ \frac{i}{f_{\pi}\, cos (\g_{\rm
S})}\Big(-Re(\Phi^I)(it_I)+Im(\Phi^I)(t_I)\Big)\Big] ~~~~,
\eqno(21) $$
and thus the generators of the group are $\{t_I,it_I\}$.

Right chiral superfield Maurer-Cartan forms $ R_{\rm I} {}^{\rm K} (\Phi)$
and left chiral superfield Maurer-Cartan forms $ L_{\rm I} {}^{\rm K} (\Phi)$
are defined by
$$ \eqalign{
U^{-1} D_{\a} U &=~ {[f_{\pi} \, cos (\g_{\rm S})]}^{-1} ( \, D_{\a}
\Phi^{\rm I} \,)~ R_{\rm I} {}^{\rm K} (\Phi) \, t_{\rm K} ~~~~,\cr
(\, D_{\a}U \,) U^{-1} &=~ {[f_{\pi} \, cos (\g_{\rm S})]}^{-1} ( \,
D_{\a} \Phi^{\rm I} \,) ~L_{\rm I} {}^{\rm K} (\Phi) \, t_{\rm K} ~~~~, }
\eqno(22) $$
and $R_{\rm I} {}^{\rm K} (\Phi) $ and $L_{\rm I}{}^{\rm K} (\Phi) $ can
be calculated as in (3). We denote the matrix inverses of $R_{\rm I}
{}^{\rm K} (\Phi) $ and $L_{\rm I} {}^{\rm K} (\Phi) $ by $ (R^{-1} )_{\rm K}
{}^{\rm I}$ and $(L^{-1} )_{\rm K} {}^{\rm I}$, respectively. Since the
multiplication of chiral superfields is closed we also observe,
$$
U^{-1} {\Bar D}_{\ad} U ~=~ 0 ~~~,~~~ U {\Bar D}_{\ad} U^{-1} ~=~ 0
{}~~~\to ~~~{\Bar D}_{\ad} R_{\rm I} {}^{\rm K} ~=~
{\Bar D}_{\ad} L_{\rm I} {}^{\rm K} ~=~ 0~~~,
$$
$${\Bar D}_{\ad} \Big[ U^{-1} D_{\a} U \,\Big] ~=~ i
{[f_{\pi} \, cos (\g_{\rm S})]}^{-1} ( \, \pa_{\un a}
\Phi^{\rm I} \,)~ R_{\rm I} {}^{\rm K} (\Phi) \, t_{\rm K} ~~~~,
$$
$${\Bar D}_{\ad} \Big[ \Big(D_{\a} U \,\Big) U^{-1} \Big] ~=~ i
{[f_{\pi} \, cos (\g_{\rm S})]}^{-1} ( \, \pa_{\un a}
\Phi^{\rm I} \,)~ L_{\rm I} {}^{\rm K} (\Phi) \, t_{\rm K} ~~~~.
\eqno(23) $$
We use a spinor notation where vector indices are denoted by
${\un a} \equiv \alpha \dot{\alpha}$.

We consider the rigid $SU_L(3) \otimes SU_R(3)$ transformations defined by
$$
\Big( U \Big)' ~=~ \exp [ - i {\Tilde \a}^{{\rm I}} t_{\rm I} \,] ~ U
{}~ \exp [ i {\a}^{{\rm I}} t_{\rm I} \,] ~~~,
\eqno(24)
 $$
with left and right transformation parameters ${\Tilde \a}^{{\rm I}}$ and
$\a^{{\rm I}}$. Note that these are {\em not} complexified transformations
(i.e. ${\Tilde \a}^{{\rm I}}$ and ${\a}^{{\rm I}}$ are real).
  Infinitesimally, using Maurer-Cartan forms, this can
be written as a variation of the chiral superfields,
$$
\d \Phi^{\rm I} ~=~  \a^{(A)} \xi^{\rm I}_{(A)} (\Phi)  ~~~,
\eqno(25)  $$
where
$$
\a^{(A)} \xi^{\rm I}_{(A)} ~\equiv ~ - \, i [ \, f_{\pi} cos (\g_{\rm
S})\, ] \, [\, {\Tilde \a}^{{\rm J}} (L^{-1} )_{\rm J} {}^{\rm I} ~-~
{\a}^{{\rm J}} (R^{-1} )_{\rm J} {}^{\rm I} ~]
{}~~~, \eqno(26)  $$
or, in finite form, as a coordinate transformation,
$$
\Big( \Phi^{\rm I}  \Big)' ~=~ K^{\rm I} (\Phi ) ~
 = ~ \exp [ \, \a^{(A)} \xi^{\rm J}_{(A)} \pa_{\rm J} \,]
\Phi^{\rm I} ~~~,~~~  \pa_{\rm J} ~\equiv ~ \pa/ \pa \Phi^{\rm J} ~~~.
\eqno(27) $$

As described in ref. \cite{jim}, we also introduce nonminimal scalar
multiplets described by complex linear superfields $\S^I$, and we split the
physical fields between the components of $\Phi$ and $\S$. In particular, for
the bosonic fields, we write (as usual, the vertical bar indicates evaluation
at $\theta = 0$)
$$ \eqalign{ {~~~~~~~~~}
\Phi^{\rm I} | &=~ {A}^{\rm I}(x) ~=~ {\cal A}^{\rm I} (x)~+~ i
\Big[ \, \Pi^{\rm I} (x) cos(\g_{\rm S} ) ~+~ \Theta^{\rm I} (x)
sin(\g_{\rm S} ) \, \Big] ~~~, \cr
\S^{\rm I} | &=~ {B}^{\rm I}(x) ~=~ {\cal B}^{\rm I}(x) ~+~ i
\Big[ \, - \Pi^{\rm I} (x) sin(\g_{\rm S} ) ~+~ \Theta^{\rm I}
(x) cos(\g_{\rm S} ) \, \Big] ~~~,}
\eqno(28) $$
in terms of two real octets of scalar spin-0 fields ${\cal A}^{\rm I}$
and ${\cal B}^{\rm I}$ as well as two real octets of pseudo-scalar spin-0
fields ${\Pi}^{\rm I}$ and ${\Theta}^{\rm I}$.
As was discussed in ref. \cite{jim}, the consistency of the model requires
$sin(2 \g_S) \ne 0$.

We postulate that the nonminimal multiplets transform under the
transformations (24) as $1$-forms (or cotangent vectors)
$$
 (\S^{\rm L})'= \Big( \pa_{\rm I} \, K^{\rm L} \Big) \S^{\rm I}
{}~~~, \eqno(29) $$
We may, however,
convert them into fields that transform as the group elements $U$
by introducing matrix valued fields $\Hat\S$:
$$
 \Hat\S\equiv \Big( \pa_{\rm I} \, U \Big) \S^{\rm I}
{}~~~. \eqno(30) $$
We emphasize that $ (\S^{\rm L})'$ and  $\Hat\S$ remain complex linear
because $K$ and $\pa_{\rm I} \,U$ are chiral and the product of a
chiral and a linear superfield is linear.

\sect{IV. 4D, N = 1 CNM Supersymmetric QCD Effective Action}

Any function of traces of {\it {appropriate}} products of $U$, $U\dg$,
$\Hat\S$, $\Hat\S\dg$ is automatically invariant under rigid $SU_L(3)
\otimes SU_R(3)$ transformations.  The minimal choice for the
$\s$-model action is (where ${\cal N}_0$, ${\cal N}_1$ are normalization
constants)
$$
{\cal S}_{\s}(\Phi, \, \S) ~=~  \int d^4 x \, d^2 \q \, d^2 {\bar \q}
~  \Big[ ~ f_{\pi}^2 {\cal N}_0 \, Tr [ \, U\dg \, U \, ] \,- \,
{\cal N}_1 Tr [ \,  \Hat\S\dg \, \Hat\S \, ] ~ \Big] ~~~~ .
\eqno(31) $$
We propose this action as the CNM chiral model
which for the group $SU(3)$ corresponds to the leading term of the
4D, N = 1 QCD effective action. Note that at this point, $\Hat\S$ is
a decoupled free field.

Having achieved this reformulation of previous results \cite{jim}
in such a way that the rigid $SU_L(3) \otimes SU_R(3)$ symmetry is
manifest in the superfield action, it is a simple step to reformulate
the Skyrme and WZNW terms described previously \cite{jim}. Consider
the real parts of the following superfield actions
$$ \eqalign{
{\cal S}_{Skyrme}(\Phi, \, \S) ~=~\, C_1 \int d^4 x \, d^2 \q \,
\eta^{{\un a} [ {\un c} } \eta^{{\un d} ] {\un b}} &{{Tr}} \Big[
(\pa_I U^{-1})(\pa_JU)(\pa_KU^{-1})(\pa_LU)\Big]\cr
&\times (\pa_{\un a} \Phi^{\rm I}\,)(\pa_{\un b} \Phi^{\rm J}\,)\,
C_{\g \, \d} \,( {\Bar D}_{{\Dot \g}} \S^{{\rm K}} \,) ( {\Bar
D}_{{\Dot \d}}\S^{{\rm L}} \, )  ~~, } \eqno(32)
$$
$$
{\cal S}_{WZNW}(\Phi, \, \S) ~=~ i \, C_0 \int d^4 x \, d^2 \q \,
\e^{{\un a}{\un b}{\un c}{\un d}} {\cal J}_{IJKL} (\pa_{\un a} \Phi^{
\rm I} \,) (\pa_{\un b} \Phi^{\rm J} \,)\, C_{\g\d}\,( {\Bar D}_{{\Dot
\g}} \S^{{\rm K}} \,) ( {\Bar D}_{{\Dot \d}}
\S^{{\rm L}} \, )~~,
$$
where ${\cal J}_{IJKL}$ is simply $\b_{ijkl}$ of equation (15) evaluated
for the chiral superfield $U$. These actions are completely equivalent
to the explicit forms that were given in  reference \cite{jim}. Written
in this way, it is manifest that these higher derivative terms are also
invariant under the $SU_L(3) \otimes SU_R(3)$ rigid transformations. We
note that the Skyrme term is free of some of the problems that plagued,
as recognized by its authors, the proposal in ref. \cite{berg} (however,
just as our model, the proposal in \cite{attempts} does not suffer from
propagating auxiliary fields).

To define the most general members of the CNM group manifold models
with rigid $SU_L(3) \otimes SU_R(3)$ symmetry, we add to (31) the integral
over full superspace of the real part of any function of traces of
appropriate products of $U$, $U\dg$, $\Hat\S$, $\Hat\S\dg$ as well as
the real part of the chiral integral of any function of traces of $U$ and
$\Bar D \Hat \S$ (with $\Bar D$'s contracted pairwise).  It can be verified
 that all of these actions
are at most quadratic in spacetime derivatives of bosonic fields. Thus,
these may be regarded as deformations of the basic $\s$-model action in (31).

To generalize to higher derivative terms with rigid $SU_L(3) \otimes
SU_R(3) $ symmetry we add to (32) the real part of a chiral integral of
a function of traces of $U$, $\Bar D \Hat \S$, and $\pa_{\un a}U$ with all
possible Lorentz contractions on the spinor indices. Some typical terms
might
be:
$$ \eqalign{ {~~~~~~}
{\cal S}_{{\rm H.\, D.}}(\Phi, \S) &=~
\int d^4 x \, d^2 \q ~ \sum_{A, p, q, k} L_k {}^A ~P^{A \, {\un c}_{1} \, ...
{\un c}_q }_{{\Dot \a}_1 \, {\Dot \b}_1 \, ... \, {\Dot \a}_p \, {\Dot \b}_p}
\times \cr
&{~~~~~~~~~~} {{Tr}} \Big[ \, {\cal J}^k(U) \prod_i^p \Big(
{\Bar D^{\Dot \a_i}\Hat\S \Big) \, \Big( \Bar D^{\Dot \b_i}
\Hat\S} \Big) \, \Big( \prod_j^q \pa_{ {\un c}_j }U \Big) ~ \Big]
~~~,~~}
\eqno(33) $$
where $ P^{A \, {\un c}_{1} \, ... {\un c}_q}_{{\Dot \a}_1 \,
{\Dot \b}_1 \, ... {\Dot \a}_p \, {\Dot \b}_p}$ denote the distinct
Lorentz invariant tensors that may be written for a fixed number of
indices. For example, in the case of the Skyrme and WZNW terms
we found
$$
P^{ {\un c} \, {\un d} ~~ {\Dot \a} \, {\Dot \b} }_{Skyrme } ~=~
\eta^{{\un a} [ {\un c} } \eta^{{\un d} ] {\un b}} C_{\a \b} ~~~, ~~~
P^{ {\un c} \, {\un d} ~~ {\Dot \a} \, {\Dot \b} }_{WZNW} ~=~
\e^{{\un a}{\un b}{\un c}{\un d}} C_{\a \b} ~~~.
\eqno(34) $$
\noindent
In (33) if $p$ is greater than one, then the superfield action when
evaluated in terms of component fields will have the property that it
possesses no purely bosonic terms. So in a sense the most interesting
actions described by (33) are those for which $p = 1$ and $q > 1$. In
particular, the simplest of these, with $p=1$, $q=2$, leads to a component
action containing the higher-derivative pion term
\cite{jim}
$$ \eqalign{
\int d^4 x &d^2 \q ~ {\cal J}_{\rm I \, J \, K \, L} (\Phi) \,
({\Bar D}^{\ad} \S^{\rm I} \, ) \, ({\Bar D}^{\Dot \b} \S^{\rm J} \, ) \,
(\pa^{\g} {}_{\ad} \Phi^{\rm K} \,) \, ( \pa_{\g \Dot \b} \Phi^{\rm L}
\,) |_{pion}{~~~~~~~~~~} \cr
 &=~ \frac 14 sin^2 (2 \g_{\rm S}) \int d^4 x\, {\cal J}_{\rm I \,
J \, K \, L} (\Pi) \, {\rm P}^{\un a \, \un b \, \un c \, \un d} \, \Big[\,
( \pa_{\un a} \Pi^{\rm I} \, ) \, ( \pa_{\un b} \Pi^{\rm J} \, )
\, (\pa_{\un c} \Pi^{\rm K} \,) \, ( \pa_{\un d} \Pi^{\rm L} \,) \,
\Big] ~~~. }\eqno(35) $$
We emphasize that all of these actions are {\em auxiliary-free}: at the
 component level the auxiliary fields {\em do not} propagate in spite of the
higher-derivative terms.

Although our formulation in terms of chiral $d^2 \theta$ integrals, based
on the simple but key observation that
${\Bar D}^{\ad} \S$ is chiral \cite{jim}, is consistent, we now know
that it is not unique. In fact, terms of the type in (33), and
specifically in (35), {\em can} be rewritten as full superspace integrals:
$$\eqalign{
\int d^4 x &d^2 \q ~ {\cal J}_{\rm I \, J \, K \, L} (\Phi) \,
({\Bar D}^{\ad} \S^{\rm I} \, ) \, ({\Bar D}^{\Dot \b} \S^{\rm J} \, ) \,
(\pa^{\g} {}_{\ad} \Phi^{\rm K} \,) \, ( \pa_{\g \Dot \b} \Phi^{\rm L}
\,) {~~~~~~~~~~} \cr
 &=~ -i\int d^4 x d^4 \q ~ {\cal J}_{\rm I \, J \, K \, L} (\Phi) \,
\S^{\rm I}  \, ({\Bar D}^{\Dot \b} \S^{\rm J} \, ) \,
(D^{\g}\Phi^{\rm K} \,) \, ( \pa_{\g \Dot \b} \Phi^{\rm L}
\,) ~~~. }\eqno(36) $$
Having rewritten this term in the action in this form, there is no longer
any obvious reason to restrict ${\cal J}$ to be  a function of chiral
superfields
 only; indeed, group
invariance is maintained even when ${\cal J}$ is a function not only of
$\Phi$, but of $\Bar \Phi, \S , \Bar\S$ as well.

One of the issues that was raised (and not answered) in the second
work of ref. \cite{jim} was whether it is possible that some auxiliary-free
higher derivative terms can arise from integrals over the full superspace
instead of chiral superspace integrals as in (33). The answer to this
is yes. It is a straight forward calculation to prove that the following
expression (which is {\em not} in general a rewriting of a chiral integral)
$$ \eqalign{
{{\cal S}'}_{{\rm H.\, D.}}(\Phi, \S) &=~
\int d^4 x \, d^2 \q \, d^2 {\bar \q} \sum_{A,k, p, q, r, s} {L'}_k {}^A
{}~{P'}^{A \, \{ {\un c}_i \} }_{\{ {\a}_i \} \, \{ {\Dot \a}_i \} }
{{Tr}} \Big[ \, {\cal J}^{A\, k\, \, \{{\rm K}_i \} \, \{ {\rm L}_i
\} }_{ \{ {\rm I}_i \} \, \{ {\rm J}_i \} }
( U \,) ~\Big] \cr
&{~~~~~~~~~~~~~~~~}\times \, \, \Big( \prod_{i= 1}^p \S^{{\rm
I}_i} \Big) \, \Big( \prod_{i= 1}^q {\Bar D}_{{\Dot \a}_i}
{\S}^{ {\rm J}_i }
\Big) \, \Big( \prod_{i= 1}^r {D}_{{\a}_i} {\Phi}^{ {\rm K}_i } \Big) \,
\Big( \prod_{i= 1}^s {\pa}_{{\un a}_i} {\Phi}^{ {\rm L}_i }
\Big) ~~~, } \eqno(37) $$
leads to an auxiliary free higher derivative action.
Thus, the analog of (13) takes the form
$$
{\cal S}_{eff}^{SUSY}(QCD) ~=~ {\cal S}_{\s} (\Phi, \, \S) ~+~
{\cal S}'_{{\rm H.\, D.}}(\Phi, \S) ~+~ {\cal S}_{WZNW} (\Phi, \S)
{}~~~, \eqno(38) $$
where the $P$-symbols in (33) can be related to those in (13) by
$$
P^{A ~ {\un c}_{1} \, {\un c}_2 \, {\Dot \a} \, {\Dot \b}} ~=~ P^{(A)
\, {\un c}_1 ,\, \b {\Dot \g}_2 ,\, \un a , \, {\g}_2 {\Dot \b}} C_{\a
\, \b} {}~~~ \eqno(39) $$
(with, as usual, ${\un b} \equiv {\beta \dot{\beta}}$ etc).  To the
same order as the Gasser-Leutwyler result we would only let $q = 2$. In
general, however, the effective action of (29) will include terms with
$q$ an arbitrary even integer.

\sect{V. The CNM Map}

Having achieved our goal of providing a supersymmetric generalization
of (13), it is apparent that more has also been accomplished.  To all
orders, the derivative terms of the pion sector of QCD effective action 
is expected to be of the form
$$
{\cal S}_{eff} (QCD) ~=~ {\cal S}_{\s} ~+~ \sum_{i = 0}^{\infty}
{}~ \Big[ ~ {\cal S}^{(i)} \Big]
{}~+~ {\cal S}_{WZNW} ~~~~, \eqno(40) $$
$$
{\cal S}^{(i)} ~=~
\int d^4 x ~ \sum_{A,B, k} L_{k} {}^{(A) (B)} \, P^{(A) \, {\un a}_{1} \, ...
{\un a}_{2 i}} \, P^{(B) \, {k}_1 ... {k}_{2 i} }_{{~~\,~~}i_{1} \,
... i_{2 i}}  \, {{Tr}} \Big[ \, {\cal J}^{A\, B \,k }_{{k}_1 ... {k}_{2 i}
} ( {{\cal Z}} \,) ~ \Big] \, \Big( \prod_{j=1}^{2 i} {\pa_{\un
a_j} \Pi^{i_j}} \Big) ~~~,
\eqno(41) $$
where for completeness, we have introduced a set of irreducible projection
operators that act on the $i$-indices.  We can define a mapping
${\cal G}_{\rm S}^C$ on higher derivative bosonic terms to the
supersymmetric ones via the following simple rules,
$$ \eqalign{
{\cal G}_{\rm S}^C : \exp [ i \frac 1{ f_{\pi}} \Pi^i t_i ] ~&\to ~~~
\exp \Big[ {{{\Phi}^{\rm I} t_{\rm I}}\over{~ f_{\pi}\, cos (
\g_{\rm S}) ~}} \Big]  ~~~, \cr
{\cal G}_{\rm S}^C : \int d^4 x  ~&\to ~~~ \int d^4 x \, \int d^2 \q
~~~, \cr
{\cal G}_{\rm S}^C : \Big( \prod_{j = 1}^q \pa_{ {\un c}_j } \Pi^{i_j}
\Big) ~&\to ~~~
C_{\g_1 \, \g_q}^{{~~}_{~}} (\, {\Bar D}_{{\Dot \g}_1} \S^{{\rm I}_1})
 \,( \, {\Bar D}_{{\Dot \g}_q } \S^{{\rm I}_q} ) \Big(
\prod_{j= 2}^{q - 1} \pa_{ {\un c}_j } \Phi^{{\rm I}_j} \Big) ~~~,
} \eqno(42) $$
for chiral actions. Similarly  non-chiral higher derivative actions are
obtained by using ${\cal G}_{\rm S}$ where
$$ \eqalign{ {~~~~~}
{\cal G}_{\rm S} : \exp [ i \frac 1{f_{\pi}} \Pi^i t_i ] ~&\to ~~~
\exp \Big[ {{{\Phi}^{\rm I} t_{\rm I}}\over{~ f_{\pi}\, cos (
\g_{\rm S}) ~}} \Big]  ~~~, \cr
{\cal G}_{\rm S} : \int d^4 x  ~&\to ~~~ \int d^4 x \, \int d^2 \q \,
\int d^2 {\bar \q}  ~~~, \cr
{\cal G}_{\rm S} : \Big( \prod_{j = 1}^q \pa_{ {\un c}_j } \Pi^{i_j}
\Big)  ~&\to ~~~  \Big( \prod_{i= 1}^P \S^{{\rm I}_i}  \Big) \,
\Big( \prod_{i= 1}^Q {\Bar D}_{{\Dot \a}_i} {\S}^{{\rm J}_i }  \Big) \,
\Big( \prod_{i= 1}^R D_{{\a}_i} \Phi^{{\rm K}_i}  \Big) \,
\Big( \prod_{i= 1}^S {\pa}_{{\un c}_i} \, \Phi^{{\rm L}_i} \Big) ~~~.}
\eqno(43) $$
where $q = P + Q + R +S$.  Thus, the most striking feature of the CNM approach
is that it is easily permits {\it {exactly the same polynomials ${\cal J}^{A\,
B\, k}$ that determine the higher derivative terms of the non-supersymmetric
QCD effective action to also determine the higher derivative terms of a 4D, 
N = 1 supersymmetric QCD effective action}}.

\sect{VI. Duality Transformation}

We now focus on the action (31) with the added term (35) rewritten as a full
superspace integral as in (36), and perform a duality transformation to a
theory with chiral and anti-chiral superfields only \cite{nonmin}.  As usual,
this is done by relaxing the constraint on the superfield to be dualized,
in this case $\Hat\S$, and imposing it by adding a Lagrange multiplier field,
in this case a chiral superfield $\chi$. Integrating out the Lagrange
multiplier reimposes the constraint and gives back the model in CNM language,
whereas integrating out the unconstrained field gives the dual model.
Explicitly, we replace (31) and (36) by
$$\eqalign {
{\cal S}_{\s}(\Phi,X) ~=~ \int d^4 x \, d^2 \q \, d^2 {\bar \q} &\left(
Tr [ U\dg U ] - Tr [ \Hat X\dg \Hat X - \chi \Hat X -\Hat X\dg\chi\dg ]
\right.\cr
- &\left. \Big[ i {\cal J}_{\rm I \, J \, K \, L} (\Phi) \,
X^{\rm I} ({\Bar D}^{\Dot \b} X^{\rm J}  )
(D^{\g}\Phi^{\rm K}) (\pa_{\g\Dot\b}\Phi^{\rm L})~+{\rm {h.c.}}
\Big]\right)~~,}
\eqno(44) $$
where $\Hat X$ is an unconstrained field replacing $\Hat\S$. Integrating out
$\Hat X$ and its conjugate gives an equation that can be solved iteratively
for them in terms of $\Phi,\chi$, their conjugates, and their derivatives.
We find
$$
\Hat X = \chi\dg + ...
\eqno(45) $$
where the remaining terms are higher order in spinor and vector derivatives.
Substituting back, to leading order, we find an action of the form proposed 
by \cite{attempts}, but with twice as many superfields.

\sect{VII. Conclusion}

We have presented a superspace formulation of  the QCD effective action
in terms of chiral and linear superfields, which can be used to map
any bosonic action into a corresponding supersymmetric one.
Within  the purely chiral superfield sector of the theory, since it is 
a 4D, N = 1 supersymmetric non-linear $\s$-model, the CNM action
defined by equation (38) obviously describes a K\" ahler manifold with 
potential\footnote{To our knowledge, this explicit form 
of the K\" ahler potential was first suggested in ref. \cite{pernici}.} 
$K(\Phi, {\Bar \Phi}) = {{Tr}} \Big[  U \, U\dg \Big] $.  More explicitly, 
it may be written as a finite sum of products of holomorphic and antiholomorphic 
functions, $K(\Phi, {\Bar \Phi}) = \sum f^{I} (\Phi) {\bar f}^I ({\Bar \Phi})$ 
for some suitable functions\footnote{These functions correspond to the ``old'' 
$\s$ and $\vec \pi$ variables in the original works on $\s$-models.} $f^I$.  
For obvious reasons, we may call this a ``group'' K\" ahler potential and 
the geometry associated with this ``group K\" ahler geometry.''

An interesting feature of the CNM formulation of the N = 1 supersymmetric
low-energy QCD effective action is the `prediction' of a new physical
constant, the non-vanishing mixing angle $\g_S$, with $sin(2 \g_S) \ne 0$.
In the CNM model of the supersymmetric low-energy QCD effective action,
requiring the presence of higher order derivative terms that contain a
component field which can be identified as the pion octet (see eq. (41))
imposes this restriction on $\g_S$. This angle is reminiscent of the
``weak mixing angle'' $\q_W$ of the Glashow-Salam-Weinberg model of the
Electroweak Interaction which is also restricted by theoretical reasons
to satisfy $sin(2 \q_W) \ne 0$ in order that the photon couple to a
{\it {purely}} vector current.


\end{document}